\documentclass[conference]{IEEEtran}
\ifCLASSINFOpdf
   \usepackage[pdftex]{graphicx}
\else
   \usepackage[dvips]{graphicx}
\fi
\usepackage{url}

\begin{document}
%
\title{A multi-protocol framework for the development of collaborative virtual environments}

\author{\IEEEauthorblockN{Luciano Argento, Angelo Furfaro}
\IEEEauthorblockA{D.I.M.E.S. \\
University of Calabria\\
P. Bucci, 41C\\
I-87036 Rende (CS), Italy\\
Email: \{l.argento, a.furfaro\}@unical.it}
}


%


\maketitle

\begin{abstract}

Collaborative virtual environments (CVEs) are used for collaboration and interaction of possibly
many participants that may be spread over large distances. 
Both commercial and freely available CVEs exist today.
Currently, CVEs are used already in a variety of different fields: gaming, business,
education, social communication, and cooperative development.
In this paper, a general framework is proposed for the development of a cooperative environment
 which is able to exploit a multi protocol network infrastructure.
The framework offers support to concerns such as communication security and  inter-protocol interoperability and let software engineers to focus on the specific business of the  
CVE under development.
To show the framework effectiveness we consider, as a case of study, the design of a reusable software layer for the development of distributed card games
  built on top of it.
 This layer is, in turn, used for the implementation of a specific card game. 


\end{abstract}


%
\IEEEpeerreviewmaketitle

\section{Introduction}
\label{sec:intro}
A collaborative virtual environment (CVE) is a space where several people, often spread over different  locations, interact with each other. The aim of these people is to share ideas
and experiences in a cooperative setting - hence the name \cite{cve}.
Recently, the term ``Web 3.0'' has been introduced to refer to the
future aspects of the Web; some groups think  that the Semantic Web
will play the role of the main new technology in this context,  while others consider CVEs as the most important advancement of the Web~\cite{OlivierPinkwart}.

Both commercial and freely available CVEs exist today.
These systems are used in a variety of different fields such as gaming, business, education
and cooperative development.
A number of companies have started exploiting CVEs, like the prominent Second Life \cite{secondlife},
for business purposes. Second Life has also been used in the educational field;
on-line lectures have been given and pedagogical usefulness of this novel medium
has been investigated by a considerable number of colleges and educational researchers~\cite{Warburton:2009}.
One of the most famous application area for 3D CVEs is gaming.
Even older games like Doom~\cite{doom}  can be considered as
an early form of CVE.

This paper proposes a general framework for the development of  cooperative virtual environments
which exploits a multi protocol network infrastructure.
The framework offers support to concerns such as communication security and
inter-protocol interoperability and let software engineers to focus on the specific
business of the CVE under development.
The framework was designed to build applications independent from
the communication technology, which means that it is possible to seamlessly add 
new communication protocols, as  new features, without affecting the remaining code. 
For instance,  an existing application, featuring only Bluetooth technology, 
can be integrated with Wi-Fi technology in a transparent way. Moreover, as stated early,
users can join the environment by using different technologies. 
To show the framework effectiveness we consider, as a case of study, the design of
reusable software layer for the development of distributed card games.
This layer is, in turn, used for the implementation of a specific card game, i.e. the Italian card game named Tressette~\cite{tressette}.

A deep study was conducted to guarantee a high degree of reusability, therefore
a great number of card games can be developed, such as Poker, Blackjack, Spades, Uno, etc.,
and the same card game can be easily  deployed on different platforms.

The rest of the paper is organized as follows. Section~\ref{sec:related} briefly surveys the related work. Section~\ref{sec:framework} presents the framework design. Section~\ref{sec:game} describes the case study.
Section~\ref{sec:conclusion} concludes the paper and outlines some possible future work. 

\section{Related work}
\label{sec:related}
Collaborative virtual environments, also referred to as \textit{collaborative
workspaces} (CW) or  {\it networked virtual environments} (NVE or NVE-VE), have been around
since the late 80s. The key areas and methodologies regarding the most pioneering CVE systems
from 1987 to 2003 are discussed in~\cite{JoslinEtAl:2004}. There exists a variety of application 
domains for CVE systems such as gaming, business and education.

A virtual environment enabling physics experiments to be cooperatively accessible via Internet, named \textit{iLabs}, has been proposed in~\cite{ScheucherEtAl:2009}. The objective of  \textit{iLabs} is to let students and educators work together in a collaborative way by using a three dimensional CVE, in the field of physics education.
Virtual reality (VR) technologies have been exploited in~\cite{Monahan:2008} to achieve a 
Collaborative Learning Environment with Virtual Reality (CLEV-R) as a sort of VR campus
where students go to learn, socialize and collaborate on-line. This research also investigates 
the use of mobile systems for e-learning.

The use of CVEs for implementing a novel business modeling approach is described in~\cite{BrownEtAl:2011}
where the authors provide a complete and usable environment for collaborative (re-)design of business 
processes by extending a number of 3-D tools.

A study on whether virtual worlds such as Second Life and World of Warcraft (a well-known MMORPG
developed by Blizzard Entertainment) can offer a basis for trade (B2C and C2C e-commerce)
and whether credible studies of e-collaboration behavior and related outcomes can be performed
on them has been conducted in~\cite{Kock:08}.

As stated in the previous section, to test the effectiveness of the  framework proposed  in this paper, a
reusable software layer for the development of distributed card games was realized
and a specific card game was developed on top of it.
Many open source projects regarding the development of a framework to build card games
can be found on the web.
AJAX Cards~\cite{ajaxcards} is a framework for the development of
single-player card games on the web by using HTML, CSS and JavaScript.
Future plans involves the inclusion of AJAX support for multiplayer games.
The Lua card game framework~\cite{luacard} provides a web page in which it is possible to develop 
multiplayer card games by uploading images for cards and scripts, written in the Lua programming language \cite{IerusalimschyEtAl:1996}, to describe the rules of the game.
None of the above referenced frameworks takes contemporaneously into account multi platform deployment,
heterogeneous communication technologies and security, which are instead the benefits inherited by using our framework for CVE development. 


\section{Framework design}
\label{sec:framework}
\begin{figure}[bt]
\centering
\includegraphics[width=0.5\columnwidth]{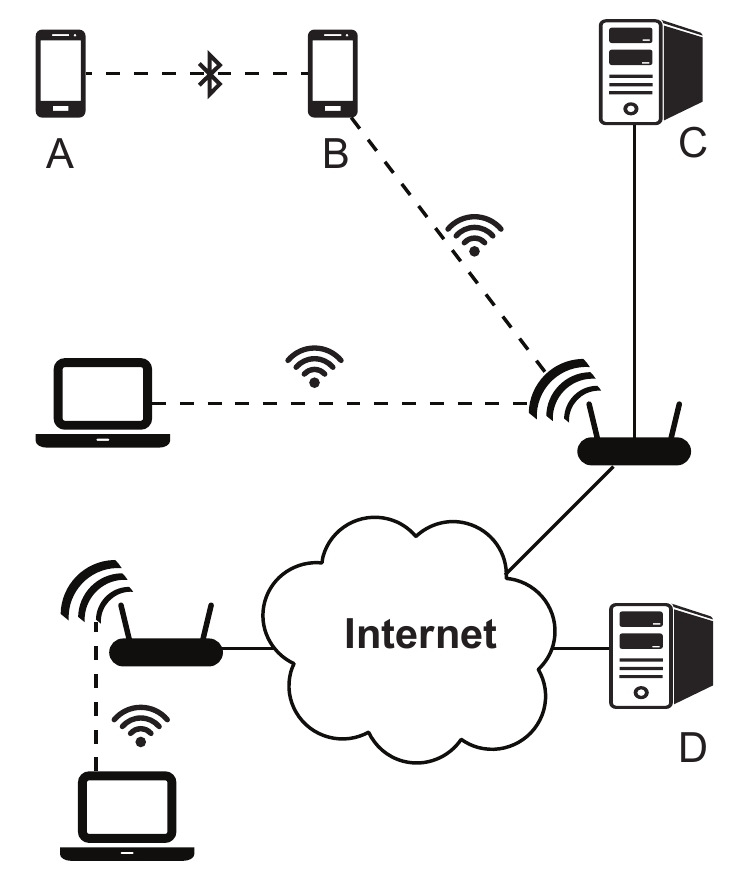}

\caption{A multi-protocol cooperative scenario}
\label{fig:scenario}
\end{figure}

This section explains in details the framework designed to
provide basic tools to build collaborative virtual environment.
The basic building block of an environment is  the \textit{Room} entity,
which represents the cooperation space on the side of each participant. 
Such entities interact among them by means of 
communication channels achieved by exploiting specific abstractions
that hide modules which handle the needed communication technologies.
A typical scenario, the framework is able to support,  is shown in Fig.~\ref{fig:scenario} where
different types of devices interact among them by using different communication media.

\begin{figure}[bt]
\includegraphics[width=\columnwidth]{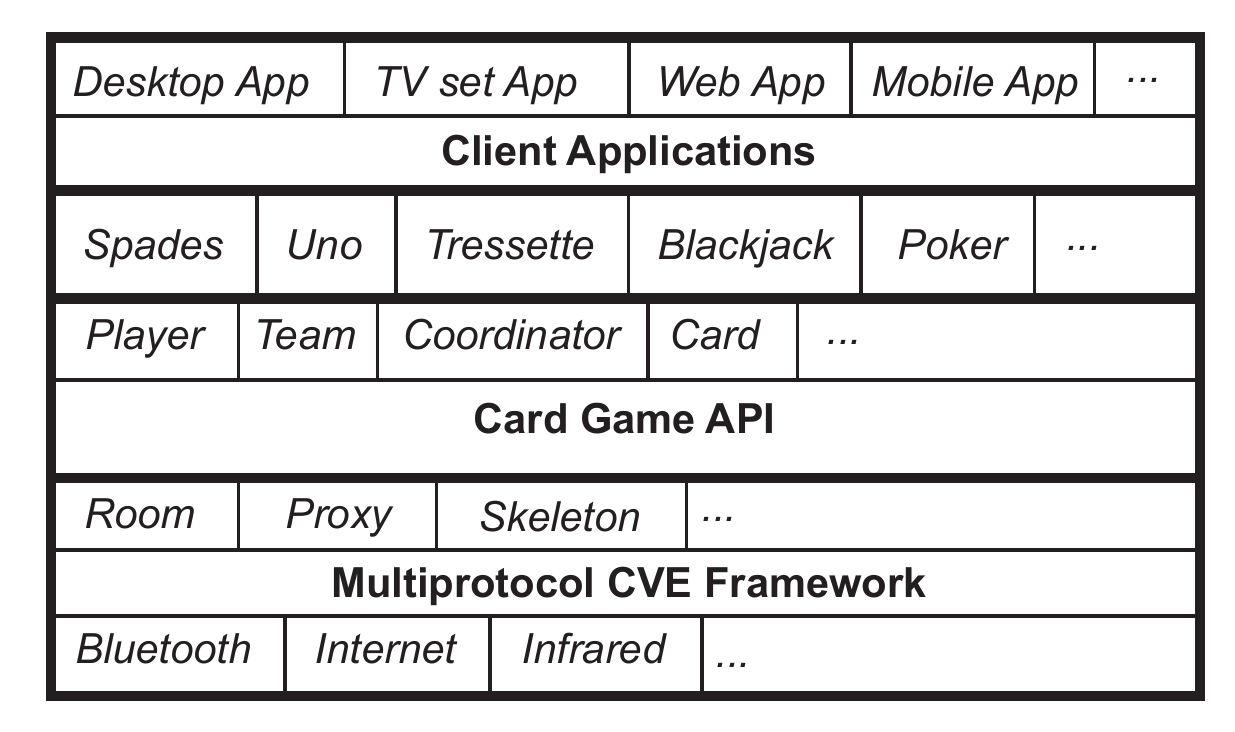}

\caption{Framework architecture}
\label{fig:fwarch}
\end{figure}

The design of the framework has been carried out by taking into account 
the following aspects:
\begin{itemize}
\item the evaluation of technologies  suitable to achieve a portable software;
\item the definition of a software architecture whose modules are as decoupled as possible
in order to guarantee a high degree of reusability, and which is able to support the deployment
of  domain-specific collaborative applications in a distributed environment;
\item the achievement of a multiprotocol network infrastructure able to shield 
clients from  details about communication technologies;  
\item the choice and analysis of a domain-specific application as a case of study in order to assess the framework effectiveness.
\end{itemize}

For the last point, we realized  a reusable software layer, for the development of distributed card games,  built on top of
the framework. Many card games were analyzed to identify the most significant and generic
entities and operations.
Subsequently, proper abstractions were defined to build a common basis for the development
of concrete card games.

The resulting architecture, shown in Fig.~\ref{fig:fwarch}, is composed of 3-layers:
\begin{itemize}
\item application clients
\item domain-specific API
\item cooperative multi-protocol framework.
\end{itemize}

The next subsections explains the study performed for this work, the solutions adopted
in order to build the mentioned layers and the way they were decoupled.

\begin{figure*}[ht]
\centering
\includegraphics[width=0.9\textwidth]{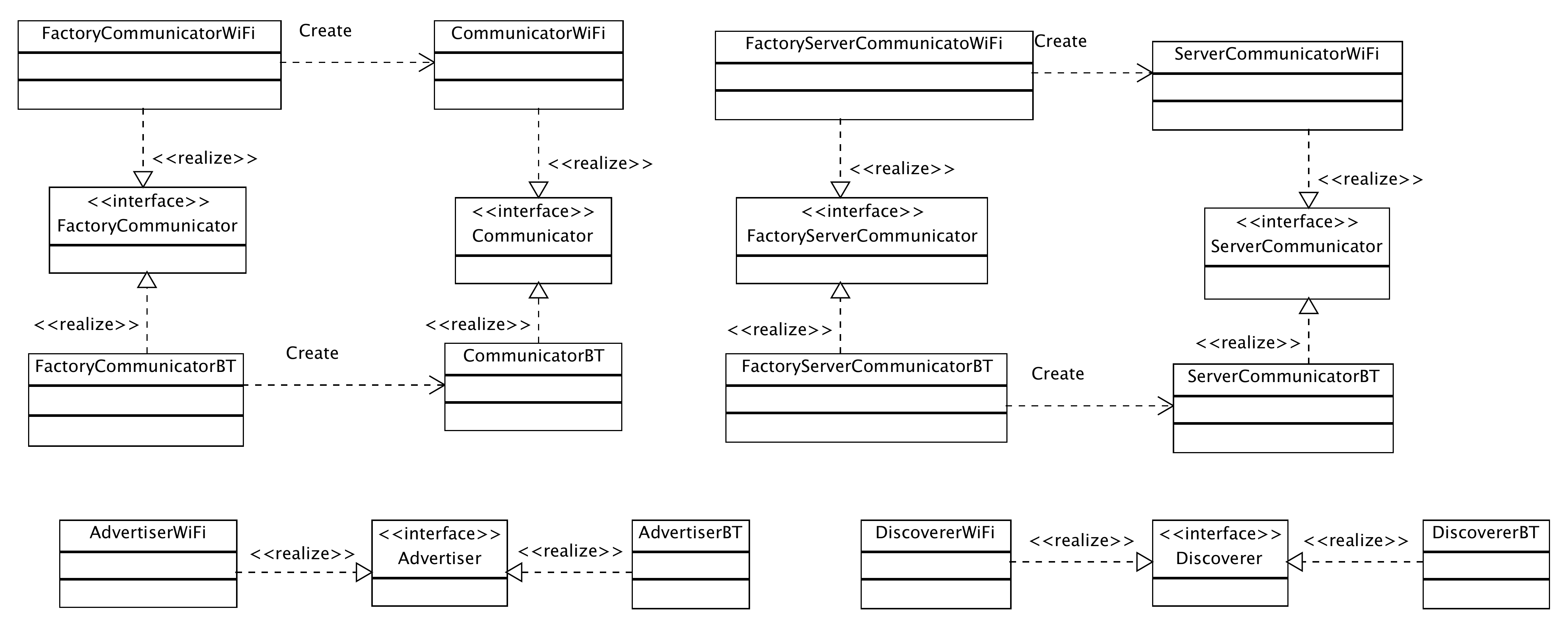}
\caption{Framework entities}
\label{fig:commdia}
\end{figure*}

\subsection{Communication}
One of the main goals of this work was to design a framework which is able to provide
a set of features suitable for the development of a multi-protocol network infrastructure.

The framework supports multiple architectural patterns for distributed communication such as
peer to peer (P2P) and client/server.
In this paper, the devices which host the component in charge of the data elaboration
will be referred to, by abuse of terminology, as servers while the others as clients.

Two distinct hierarchies were designed to handle all the aspects regarding
the communication between devices.
The highest level   abstractions of the mentioned hierarchies are represented by the following
two interfaces: \textit{ServerCommunicator} and \textit{Communicator}.

The first interface provides services thought to create a communication channel
while the second one deals with the handling of the connection.
Only these two interfaces  are accessible from client code,
the implementations of the concrete classes cannot be seen.
The design pattern \textit{Factory}~\cite{GoF:1995} has been used in order to create concrete instances.
This allows the entities which need to communicate with
a remote object to transparently use different communication protocols just by resorting to the suitable concrete
factory object. 


\subsubsection{Devices scanning}
To join a virtual environment, advertising and discovering operations are needed.
Two interfaces, \textit{Discoverer} and \textit{Advertiser}, were respectively devised in order to let a player search for an existing environment or create a new one.

To build a multi-protocol network infrastructure, different implementations of these two interfaces
can be exploited. For instance, if an application provides Bluetooth and
Wi-Fi connection, a user could advertise the cooperation environment, he previously created,
by using both Wi-Fi-based and
Bluetooth-based implementations of the \textit{Advertiser} interface, simultaneously.
On the other hand, a user that wants to join an environment will use only
the needed implementation of \textit{Discoverer}.
During the registration phase the server creates a  channel with each client
based on the chosen communication technology. These details are hidden by
the \textit{ServerCommunicator} and \textit{Communicator} interfaces
so there is no need to concern with these aspects during the development
of the domain-specific cooperation logic.

\subsection{The environment}
This section describes the modules designed to build the environment in which the cooperation takes place.
Specifically, the \textit{Room} entity has been devised in order to achieve such a goal.
 It is in charge of handling basic operations such as participants' registration,  opening
and closing of the cooperative sessions  and provides services which let a participant
to join the environment, send and receive data.
To account of the participant role, the following specific interfaces were defined: \textit{ServerRoom} and 
 \textit{ClientRoom}.
\textit{Room}  data are hosted on the side of the participant playing the server role, so it is necessary to 
make them remotely accessible to the other participants.
The \textit{Remote Proxy} design pattern offers a solution to this problem by hiding communication details and
making data to appear as they were locally available.  
Actually, the goal of the  \textit{Remote Proxy} pattern is to give a local representation for an object
that \textit{lives} in a different address space~\cite{GoF:1995}.

The \textit{proxy} entity implements the \textit{ClientRoom} interface
on the client device while on the server a \textit{skeleton} entity is created.
The hierarchies previously introduced are exploited to obtain a modular 
separation that leads to  a system which can be naturally evolvable, i.e.  it is possible
to easily integrate  new communication technologies without requiring modification of the existing code.


\subsection{The client applications}
One of the framework requirements is the capability
of binding any client application to the cooperative environment built on top of it, 
regardless of the application nature.
This decoupling  has been obtained by resorting to the \textit{Observer} pattern~\cite{GoF:1995}. 
The object representing  the cooperation room, maintains the reference to
a list of observers whose aim is to receive notification data about the status of the environment
and update the respective client application.
For instance let us suppose there are four participants $A$, $B$, $C$ and $D$.
With reference to Fig.~\ref{fig:scenario}, $A$ uses  an Android app on a mobile phone connected, via bluetooth, to $B$, which also uses an Android app and is in turn connected, via Wi-Fi to a LAN, $C$ is a desktop client connected to same LAN and $D$
a web application which runs over the Internet and it is reachable from the LAN through a router. 
Each of these applications has to register itself as observer of the room
and be able to properly handle  data received from the environment.

\subsection{Session handling}
While a cooperation session is active, it may happen that a user disconnects, e.g. because a connection problem
or because he needs to temporarily leave.
During the registration phase, the host, into which the room resides, generates a token for each participant.
This token is used to handle the session accordingly to an application specific policy.

For example in a game environment, these disconnection events could last for few seconds, thereby if they are handled properly
it is possible to avoid ending the game in unexpected way.
When a player leaves temporarily a game, he can rejoin the environment by using the token
previously obtained.
The main benefit of the use of the tokens is that only players which
originally joined the environment can rejoin after a disconnection.
Obviously a timeout can be set for the session by the server, so if a user fails to rejoin
before the time runs out, the session, will be terminated. Tokens can also carry identification
information, e.g. digitally signed messages, allowing users' authentication.

\section{Case study}
\label{sec:game}
\begin{figure}[bt]

\includegraphics[width=\columnwidth]{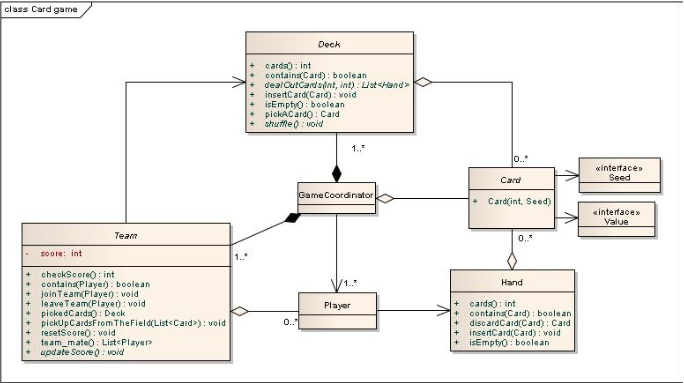}
\caption{Card game entities}
\label{fig:cardgame}
\end{figure}

As a case of study, the design of a reusable software layer for the development of distributed
card games was built on top of the framework.
In order to build such a layer it was necessary to study in depth the main characteristics
of existing card games and figure out what elements can be considered
as a basis on which to build a new card game.
The following aspects were analyzed, during the study:
\begin{itemize}
\item the possible set of values associated with the cards;
\item the set of seeds associated with the cards;
\item order relationship among cards;
\item number of players and possible partnerships;
\item actions available for a player during his turn;
\item actions available for a player during the turn of another one;
\item the interaction among players;
\item score computation;
\item card picking;
\item card distribution;
\item how the next turn is chosen;
\item the conditions by which a player or a group of players is declared as the winner.
\end{itemize}
The fundamental entities and relations of a generic card game were identified, as a result of the study,
and are shown in the  class diagram depicted in Fig.~\ref{fig:cardgame}.

\subsection{Architecture for a generic card game}
A certain number of abstraction were designed in order to capture the essential elements
which characterize a generic card game.
The abstract class \textit{Card} can be used to model cards of different games, it implements
the \textit{Comparable} interface so that a comparison is possible among the cards.
\textit{Seed} and \textit{Value} interfaces are abstractions which describe
the corresponding two features of the card concept.
The entities deck and team are respectively represented by the abstract classes \textit{Deck} and
\textit{Team}. These classes provide some services which are
common to every card game. 
Other operations, i.e. the way  the cards are 
distributed,  depend on the nature of the game: in Poker hands are of five cards,  
in Tressette with four players, see section~\ref{sec:tressette}, ten cards are 
given, at game begin, to each player.

The operations needed to handle the hand of a player, the set of cards distributed to every player,
are generic enough to suit a generic card game well, so the concrete class \textit{Hand} has been purposely devised.

There are two more entities that were identified during the mentioned study:
the player and the game coordinator. The first  serves the purpose of letting
a user interact with the game performing actions like
checking the hand, playing one or more cards and so on.
The second  has different tasks to accomplish, such as  elaborating and communicating
information about the current status of the game and coordinating the actions of all players.
The player and the game coordinator will not necessarily be located
on the same machine.
The framework has been thought to build a distributed system in which
each device can either adopt the role of a server or the role of a client, so
the entity player will be assigned to every user while the entity game coordinator
only to the user whose device act as a server. It is also possible to implement a centralized 
coordination approach where the coordinator is not tied to any player and reside on a different machine.

The game coordinator knows the players who participate in the game so
a communication between objects on different machines will take place.
The \textit{Player} interface and the abstract class called \textit{AbstractPlayer}
are exploited to hide the true nature of a player, so the game coordinator
will never know whether it is communicating with a remote object,
whether it is interacting with the local user or whether it is interacting with and intelligent artifact.

\begin{figure}[tb]
\centering
\includegraphics[width=0.85\columnwidth]{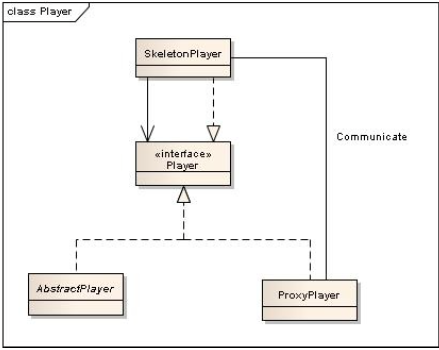}
\caption{Player entities}
\label{fig:PlayerClassDia}
\end{figure}

The \textit{Proxy} pattern~\cite{GoF:1995} was exploited so that a transparent communication channel can be created
between the player and the game coordinator (see Fig.~\ref{fig:PlayerClassDia}).
The \textit{ProxyPlayer} class implements the \textit{Player} interface and
forwards all the requests, received by the game coordinator, to its associated client.
On the client machine the \textit{SkeletonPlayer} module, another class that implements the
\textit{Player} interface and knows the real player, receives the forwarded requests.

The communication between a server and a client is bidirectional, so
the player needs a way to communicate to the game coordinator.
The game coordinator carries out a number of operations; the whole set of operations
can be split up in subsets, each of them deals with a specific aspect of the system.
In the light of the previous consideration, three interfaces were designed:
\begin{itemize}
\item \textit{GameCoordinatorElaboration};
\item \textit{GameCoordinatorCommunication};
\item \textit{GameCoordinatorRegistration}.
\end{itemize}
The first interface defines operations regarding the server, while the second and
the third ones were thought for the client.
The \textit{Remote Proxy} pattern was used once again to let a player
communicate to the game coordinator. This time the proxy and the correspondent
skeleton implement the \textit{GameCoordinatorCommunication} and \textit{GameCoordinatorRegistration}
interfaces.
To gather the operations of the two interfaces mentioned previously and to hide
the implementation details of the proxy and the skeleton, the following
two interfaces were created:
\begin{itemize}
\item \textit{ProxyGameCoordinator};
\item \textit{SkeletonGameCoordinator}.
\end{itemize}
It was also created the \textit{GameCoordinator} abstract class which implements
the \textit{GameCoordinatorElaboration}, \textit{GameCoordinatorCommunication}
and \textit{GameCoordinatorRegistration} interfaces, to provide a default implementation
for a portion of the whole set of operations and a module to handle
properly the entity game coordinator on the server.
The player and the game coordinator, just like the room entity, use the services
provided by the \textit{ServerCommunicator} and \textit{Communicator} interfaces (see Fig.~\ref{fig:commdia}).

\subsection{Security aspects}
During the design phase the existence of fake client application was taken into account.
Therefore the system was designed so that the server checks the validity of the data
received by each client. Whenever a fake client tries to play an unexpected card  
the system would be able to recognize this anomaly.
If an anomaly is detected a notification will be sent to each client to communicate
the presence of a fake client and subsequently the game would be ended.

\subsection{Tressette: a specific card game}
\label{sec:tressette}
Tressette, an Italian card game, was built on top of the framework. The application
was thought for Android and personal computer devices; the Java programming language
was used in order to guarantee the portability of the software and the Bluetooth and
the Socket technologies were chosen as communication technologies.

Tressette is played with a standard Italian 40-card deck and the cards are ranked as
follows from highest to lowest: 3-2-Ace-King-Knight-Knave and then all the remaining
cards in numerical order from 7 down to 4. The game may be played with four players
playing in two partnerships, or in heads-up play. In either case, ten cards are dealt to each player.
In the developed application the players play only in two partnerships.
The object of the game is to score as many points as possible until a score of 21
is achieved. Players must follow suit unless that suit does not remain in their hand,
and players must show the card they pick up off the card pile to their opponent.
More information about the Tressette card game can be found in~\cite{tressette}.

\begin{figure*}[tb]
\centering
\includegraphics[width=0.85\textwidth]{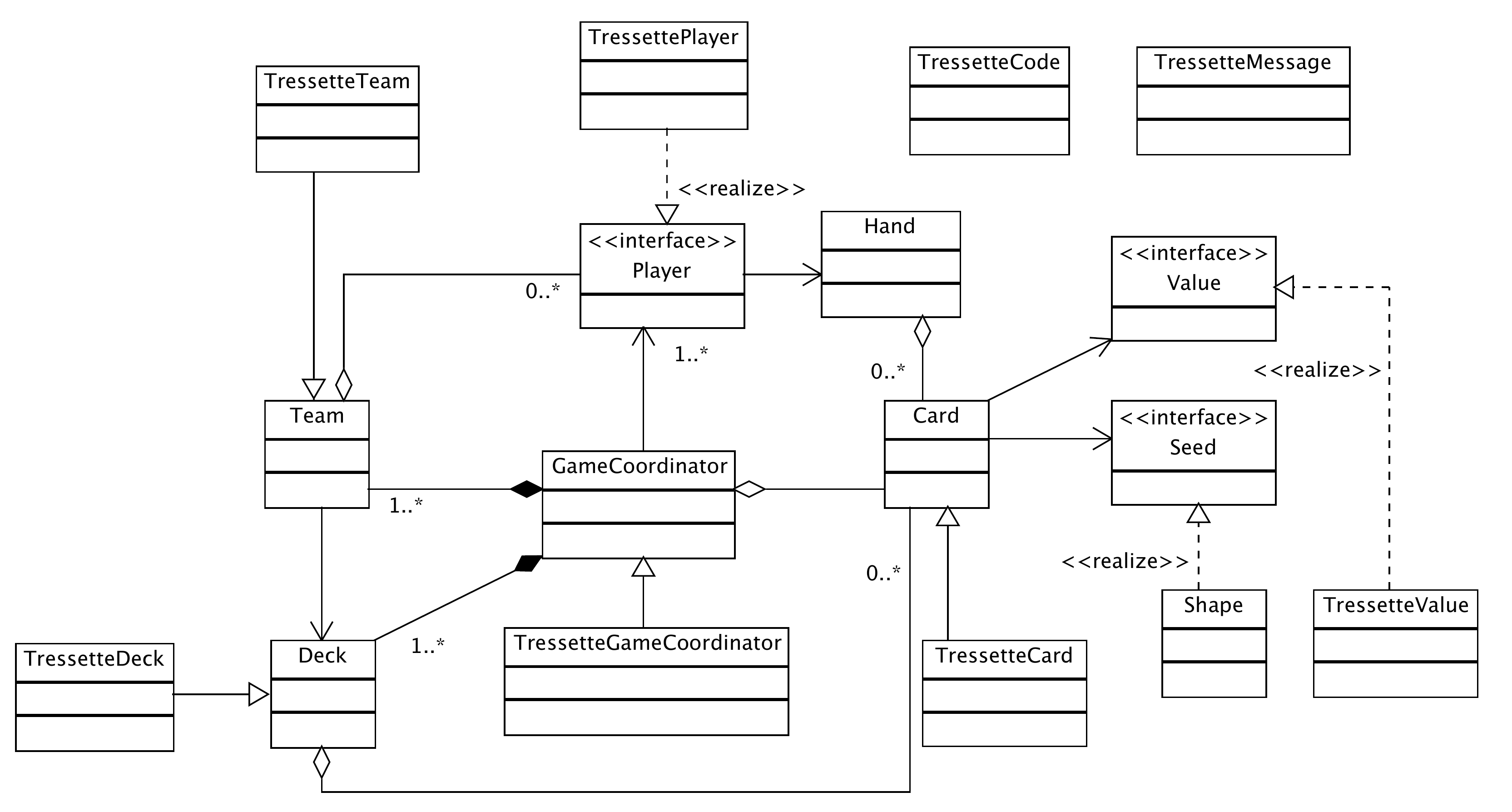}
\caption{Tressette game entities}
\label{fig:37ClassDia}
\end{figure*}

The entities that have been introduced for implementing the Tressette game are reported in the class diagram 
depicted in Fig.~\ref{fig:37ClassDia}. 
There was no need to override any methods of the \textit{Hand} class, its services
resulted generic enough for the application.
\textit{TressetteCard} is a concrete class which extends \textit{Card}, it introduces an operation
to mark a card as playable or not: in this game a card can be played only when
specific conditions hold, see~\cite{tressette} for more details.
The Enum \textit{Shape} represents the concept of seed, so it implements
the \textit{Seed} interface, while the concrete class called \textit{TressetteValue} realises
the \textit{Value} interface.
\textit{TressetteTeam} extends \textit{Team}, it keeps track of the game score. Analogously \textit{TressetteDeck}
extends \textit{Deck} with game-specific deck handling aspects.  
In the method \textit{elaboration} of the \textit{TressetteGameCoordinator} class lies the core
of the logic of the game, here the turns and scorers are updated and the winners
are proclaimed.
\textit{TressettePlayer} provides specific operations for Tressette, like the \textit{TressetteCard}
class does.
It was necessary to define two more interfaces to handle specific aspects of the game: \textit{TressetteCode} and
 \textit{TressetteMessage}. The first interface serves the purpose of exchanging messages between the room and
a client, while the second one defines all the statements available for a user.

\section{Conclusions and future work}
\label{sec:conclusion}
We presented a modular framework for the development of domain specific CVEs. 
One of the main feature of our proposal is the support for a transparent and
seamless multi protocol interaction among participants. The framework is not
tied to any particular architectural distributed communication pattern and can 
be exploited to support any of them.
The effectiveness of the framework has been experimented by implementing on top of
it a reusable software layer for the development of distributed card games.
Future research directions include others practical experimentation of the framework
in more heterogeneous distributed settings, e.g. involving web-services~\cite{DeAntonellisEtAl:2003} and cloud based applications~\cite{MalikeEtAl:2012}.    
The proposed framework can be used for building collaborative environments in many fields.
For example, existing  e-learning platforms based on CVE (e.g. \cite{BourasEtAl:2001}) could be extended in order to gain the multi-protocol features we have introduced in the framework. 
\IEEEtriggeratref{7}


\bibliographystyle{IEEEtran}
\bibliography{IEEEabrv,references}
%



\end{document}